\documentclass[sigconf,nonacm]{acmart}
\usepackage{siunitx}
\usepackage{booktabs}
\usepackage{multirow}
\AtBeginDocument{%
  }

\begin{document}

\title{Towards Trustworthy Multimodal Recommendation}

\author{Zixuan Li}
\affiliation{%
  \institution{Renmin University of China}
  \city{Haidian Qu}
  \state{Beijing Shi}
  \country{China}}



% **摘要**

% 多模态推荐的最新进展证明了将视觉和文本内容融入协同过滤的有效性。然而，实际部署引发了一个日益重要但尚未得到充分探索的问题：**可信性**。在现代电子商务平台上，多模态内容可能是误导性或不可靠的（例如，视觉不一致的商品图像或“标题党”标题），这会向多模态表示中注入不可信的信号，使得现有的多模态推荐系统在模态受损时变得脆弱。

% 在这项工作中，我们从方法和分析两个角度出发，朝着可信多模态推荐迈出了一步。首先，我们提出了一种即插即用的**模态级纠偏**组件，通过学习商品与多模态特征之间的软对应关系，来减轻不可信模态特征的影响。利用轻量级投影和基于 Sinkhorn 的软匹配，该纠偏组件在抑制失配模态信号的同时保留了语义一致的信息，并且可以在不修改架构的情况下集成到现有的多模态推荐模型中。

% 其次，针对噪声协同信号下的**交互级可信性**，我们提出了两个实际见解：(i) 在噪声环境下，训练集中的伪交互对性能的影响（提升或损害）取决于先验与信号的对齐情况；(ii) 传播图中的伪边也可能提升或损害鲁棒性，因为消息传递可能会放大这种失配。

% 这些发现强调，在交互不可信的情况下，由协同先验驱动的图增强策略并不总是有效；实现可信的多模态推荐需要对内容和交互信号进行仔细处理。在多个数据集和骨干模型上、不同受损水平下进行的大量实验证明，模态纠偏提升了系统的鲁棒性，并验证了上述交互级的观察结果。匿名代码见：https://anonymous.4open.science/r/TMR-3127。

\begin{abstract}
Recent advances in multimodal recommendation have demonstrated the effectiveness of incorporating visual and textual content into collaborative filtering. However, real-world deployments raise an increasingly important yet underexplored issue: trustworthiness. On modern e-commerce platforms, multimodal content can be misleading or unreliable (e.g., visually inconsistent product images or click-bait titles), injecting untrustworthy signals into multimodal representations and making existing multimodal recommenders brittle under modality corruption. In this work, we take a step towards trustworthy multimodal recommendation from both a method and an analysis perspective. First, we propose a plug-and-play modality-level rectification component that mitigates untrustworthy modality features by learning soft correspondences between items and multimodal features. Using lightweight projections and Sinkhorn-based soft matching, the rectification suppresses mismatched modality signals while preserving semantically consistent information, and can be integrated into existing multimodal recommenders without architectural modifications. 
Second, we present two practical insights on interaction-level trustworthiness under noisy collaborative signals: (i) training-set pseudo interactions can help or hurt performance under noise depending on prior--signal alignment; and (ii) propagation-graph pseudo edges can also help or hurt robustness, as message passing may amplify misalignment.
These findings highlight that graph enrichment strategies driven by collaborative priors are not universally beneficial under untrustworthy interactions, and that trustworthy multimodal recommendation requires careful handling of both content and interaction signals. Extensive experiments on multiple datasets and backbones under varying corruption levels demonstrate improved robustness from modality rectification and validate the above interaction-level observations.
\end{abstract}

\begin{CCSXML}
<ccs2012>
   <concept>
       <concept_id>10002951.10003317.10003347.10003350</concept_id>
       <concept_desc>Information systems~Recommender systems</concept_desc>
       <concept_significance>500</concept_significance>
       </concept>
   <concept>
       <concept_id>10002951.10003317.10003371.10003386</concept_id>
       <concept_desc>Information systems~Multimedia and multimodal retrieval</concept_desc>
       <concept_significance>500</concept_significance>
       </concept>
 </ccs2012>
\end{CCSXML}

\ccsdesc[500]{Information systems~Recommender systems}
\ccsdesc[500]{Information systems~Multimedia and multimodal retrieval}

\keywords{Multimodal Recommendation, Trustworthy Recommendation, Graph Refinement, Graph Neural Networks, Multimodal Consistency}
\maketitle

\section{Introduction}

% 在信息爆炸时代，推荐系统已成为缓解信息过载的关键技术，通过分析历史用户行为提供个性化建议。然而，传统的协同过滤算法仅依赖于用户与物品的交互历史，往往面临严重的数据稀疏问题，从而限制了推荐的准确性和覆盖率。为了应对这一局限性，多模态推荐系统应运而生，通过整合丰富的物品内容（如产品图像和文本描述）来增强用户偏好建模，特别是针对稀疏和冷启动物品。

% 沿着这一方向，多模态推荐已从简单的特征融合演进到基于图的建模。早期方法将视觉或文本特征投影到低维空间，并将其与ID嵌入相结合。较新的方法则采用轻量级图消息传递来捕捉高阶协同信号，并进一步通过相似度图利用模态引导的物品关系。也有一些研究通过交互图优化或模态噪声抑制来研究鲁棒性。尽管取得了这些进展，但一个常见且隐含的假设是多模态内容是可靠的，且观察到的交互能真实反映用户偏好。在现实世界的平台中，这一假设正日益受到挑战：物品内容可能是误导性但看似合理的（如图像与文本不匹配、标题党），隐式反馈也可能包含不可信的交互（如误点或受曝光驱动的行为）。我们将这一实际挑战称为多模态推荐中的可信性问题：用于偏好学习的多模态信号和交互边的可靠性可能存在巨大差异，这种不一致性可能会被图传播放大，甚至干扰常规的训练和评估实践，而不仅仅表现为随机的特征扰动。

In the era of information explosion, Recommender Systems (RS) have become a pivotal technology for alleviating information overload by analyzing historical user behaviors to provide personalized suggestions \cite{grouplens, nextreco,ecomm,handbook}. However, traditional Collaborative Filtering (CF) algorithms, which rely exclusively on user--item interaction history, often suffer from severe data sparsity, limiting recommendation accuracy and coverage\cite{bpr,coldstart,coldstart2,sparce}. To address this limitation, \textbf{Multimodal Recommender Systems (MRSs)} have emerged by integrating rich item content---such as product images and textual descriptions---to enhance user preference modeling, especially for sparse and cold-start items\cite{bm3,grcn,freedom,slmrec,dragon}.

Along this line, multimodal recommendation has evolved from simple feature fusion to graph-based modeling. Early methods (e.g., VBPR \cite{vbpr}, DeepStyle \cite{deepstyle}) project visual or textual features into low-dimensional spaces and combine them with ID embeddings. More recent approaches employ lightweight graph message passing (e.g., LightGCN \cite{lightgcn}) to capture high-order collaborative signals, and further exploit modality-induced item relations via similarity graphs (e.g., LATTICE \cite{lattice}). Several works also study robustness via interaction-graph refinement or modality-noise suppression (e.g., FREEDOM \cite{freedom}, MGCN \cite{mgcn}, SMORE \cite{smore}, PGL \cite{pgl}). Despite these advances, a common but implicit assumption is that multimodal content is reliable and that observed interactions faithfully reflect user preference. This assumption is increasingly violated in real-world platforms, where item content can be misleading yet plausible (e.g., image--text mismatch, click-bait titles) and implicit feedback can include unreliable interactions (e.g., misclicks or exposure-driven behaviors). We refer to this practical challenge as the \textbf{trustworthiness} issue in multimodal recommendation: the reliability of multimodal signals and interaction edges for preference learning may vary substantially, and such inconsistencies can be amplified by graph propagation and even confound common training/evaluation practices, rather than behaving as random feature perturbations.

\begin{figure}[t]
    \centering
    \includegraphics[width=\linewidth]{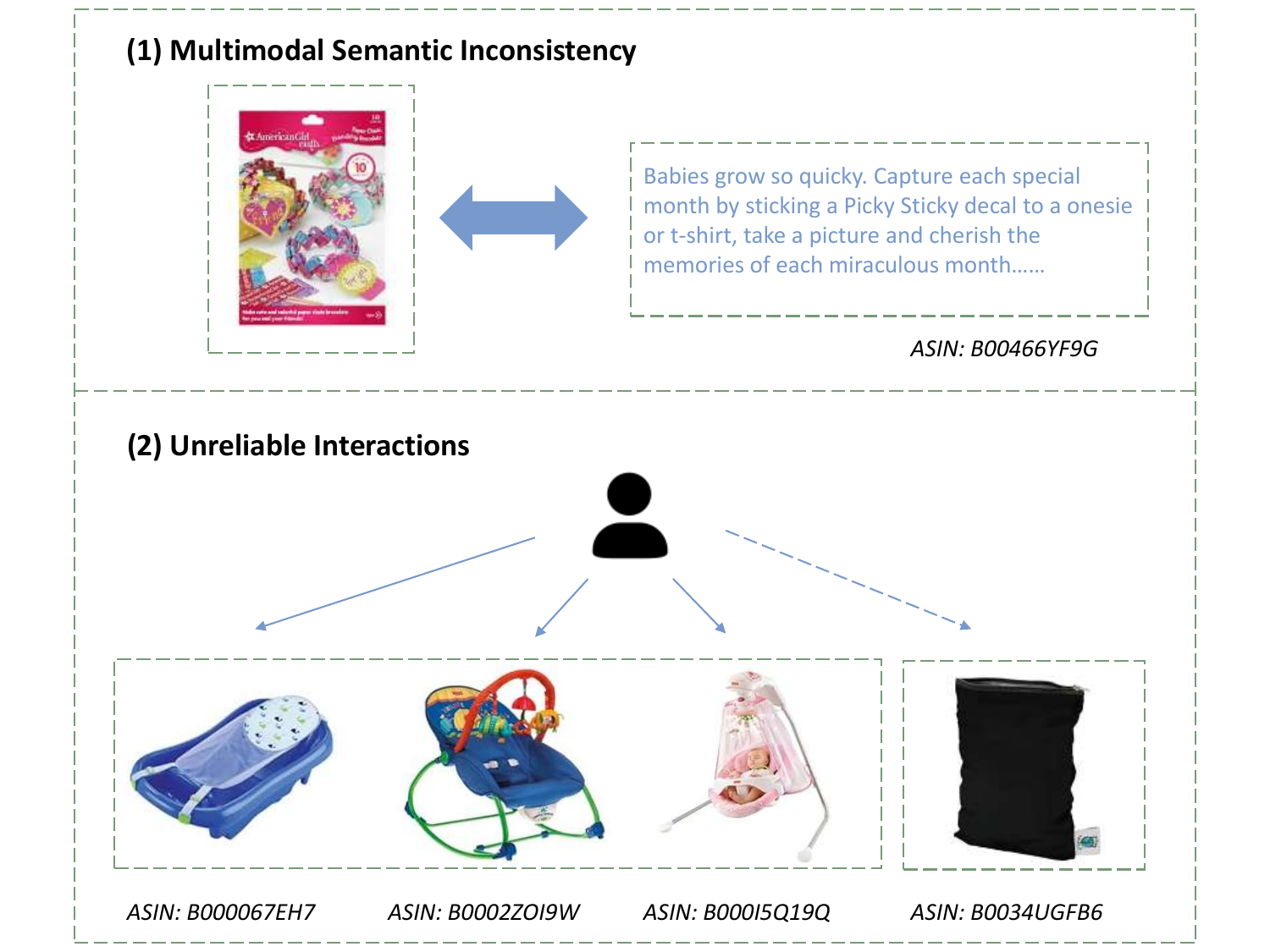}
    \caption{Illustration of untrustworthy signals in multimodal recommendation. 
    (1) Multimodal semantic inconsistency between product images and textual descriptions. 
    (2) Unreliable user--item interactions, where weakly related but plausible interactions introduce low-confidence edges that may pollute message passing.}
    \label{fig:case}
\end{figure}

% 为了说明这个问题，我们展示了一个基于Amazon Baby数据集的案例研究（图1）。首先，产品图像显示的是一个儿童友谊手链手工包，而文本描述讨论的是婴儿里程碑连体衣贴纸，导致视觉和文本模态之间存在严重的语义不一致。这种看似合理但具有误导性的内容在现实世界的平台中很常见，当模型将所有模态视为同等可靠时，可能会引发错误的跨模态关联。其次，同一图中显示的用户交互集以母婴产品为主，其功能和使用场景高度一致（例如浴盆、婴儿摇椅和摇篮秋千），同时也包含了一个与用户主要兴趣语义相关性较弱的产品（如一个可重复使用的湿袋）。这表明在隐式反馈日志中存在不可信的交互：相关性较弱但看起来合理的点击或购买可能会在消息传递过程中被无差别地传播，从而污染协同信号。这些观察结果促使我们从两个角度研究可信的多模态推荐：（1）内容信任，即在特征层面纠正具有误导性的模态信号；（2）交互信任，即理解基于图的增强何时可能会放大不可靠的协同关系。

% 为此，我们提出了一个简单且即插即用的模态级纠偏组件，它可以集成到现有的多模态推荐系统中，而无需修改其架构。其核心思想是学习物品与多模态特征之间的软对应关系，并通过强调语义一致的信号、同时降低失配信号的权重来纠正模态表示。我们利用轻量级投影和基于Sinkhorn的软匹配程序来实现这一构想，在模态失配的情况下提供了一种有原则且稳定的对应关系近似。

% 除了上述组件外，我们还进行了交互级可信性分析，并得出了对可信多模态推荐至关重要的两个实际见解。

% （见解1）在协同信号受损的情况下，向训练集中注入伪交互（例如来自协同先验或数据增强）是一把双刃剑：当注入的先验与底层协同信号一致时，它可能会提高性能；但当二者失配（misaligned）时，则可能降低系统的鲁棒性，即便在严格排除验证和测试数据的情况下也是如此。

% （见解2）在噪声交互下，仅在传播图中注入此类伪边同样可能提升或损害性能；特别是，图消息传递可能会放大失配的关系，从而导致鲁棒性下降。综合来看，这些发现表明对待生成的交互应当谨慎：它们可能会悄然改变基于图的推荐系统的有效学习动态，当协同信号不可信时，其与图传播的整合应当受到严格控制。

% 本工作的主要贡献总结如下：

% 多模态推荐中的可信性。我们强调并形式化了多模态推荐系统中由误导性多模态内容和不可靠隐式交互引起的可信性挑战，并得到了实证案例研究和受控受损设置的支持。

% 模态级纠偏。我们提出了一个即插即用的纠偏组件，通过轻量级投影和基于 Sinkhorn 的软匹配学习物品与模态之间的软对应关系，有效抑制失配的模态信号，同时保留一致的语义。

% 可信学习的交互级见解。我们证明了在协同信号受损的情况下，训练时的伪交互和传播时的伪边对鲁棒性的影响（提升或损害）取决于先验与信号的对齐情况。这强调了当交互不可信时，图增强并不总是受益的。

% 有效性和通用性。在多个真实数据集和骨干模型上进行的大量实验证明，所提出的纠偏方案提高了模态受损情况下的鲁棒性，并在噪声协同设置下验证了上述交互级见解。

To illustrate the problem, we present a case study based on the Amazon Baby dataset (Figure~\ref{fig:case}). First, the product image shows a children’s friendship bracelet craft kit, while the textual description discusses baby milestone onesie stickers, resulting in a severe semantic inconsistency between visual and textual modalities. Such misleading-yet-plausible content is common on real-world platforms and can induce erroneous cross-modal correlations when models treat all modalities as equally reliable. Second, the same figure shows a user interaction set dominated by maternal and infant products with highly consistent functionality and usage scenarios (e.g., bathtub, infant rocker, and cradle swing), while also containing a product (e.g., a reusable wet bag) that exhibits weak semantic relevance to the user’s primary interests. This suggests that untrustworthy interactions exist in implicit-feedback logs: weakly related but plausible clicks/purchases can be indiscriminately propagated during message passing, thereby polluting collaborative signals. These observations motivate studying trustworthy multimodal recommendation from two perspectives: (i) content trust, i.e., rectifying misleading modality signals at the feature level, and (ii) interaction trust, i.e., understanding when graph-based enhancement may amplify unreliable collaborative relations.

To this end, we propose a simple, plug-and-play modality-level rectification component that can be integrated into existing multimodal recommenders without architectural changes. The key idea is to learn soft correspondences between items and multimodal features, emphasizing semantically consistent signals while down-weighting mismatched ones. We instantiate it with a lightweight projection and Sinkhorn-based soft matching, yielding a principled and stable correspondence approximation under modality mismatch.

Beyond this component, we conduct an interaction-level trustworthiness analysis and derive two practical insights. (\textbf{Insight~1}) Under corrupted collaborative signals, injecting pseudo interactions (e.g., from collaborative priors or augmentation) into the training set is a double-edged sword: it can help when injected priors align with true patterns, but degrade robustness when misaligned, even with strict validation/test exclusion. (\textbf{Insight~2}) Injecting pseudo edges only into the propagation graph can likewise help or hurt; in particular, message passing may amplify misaligned relations and reduce robustness. Overall, generated interactions should be used cautiously, and their integration with graph propagation should be carefully controlled when interaction trustworthiness is low.

The primary contributions are summarized as follows:
\begin{itemize}
\item \textbf{Trustworthiness in Multimodal Recommendation}. We identify and formulate trustworthiness failures in MRSs arising from misleading multimodal content and unreliable implicit interactions, and instantiate them with reproducible controlled stress tests.
\item \textbf{Modality-level Rectification}. We propose a plug-and-play module that learns item--modality soft correspondences via lightweight projection and Sinkhorn-based matching, suppressing mismatched signals while preserving consistent semantics.
\item \textbf{Interaction-level Insights}. We show that training-time pseudo interactions and propagation-time pseudo edges are not universally beneficial under corrupted collaborative signals: they can help or harm depending on prior--signal alignment, and propagation may amplify misaligned relations.
\item \textbf{Effectiveness and Generality}. Experiments across multiple datasets and backbones demonstrate improved robustness under modality corruption and validate the above interaction-level insights under noisy interactions.
\end{itemize}

% 多模态推荐
% 多模态推荐通过引入图像和文本等物品端内容来增强协同信号，旨在缓解交互稀疏性并提高偏好估计的准确性。早期的多模态推荐系统通常将预训练编码器提取的视觉（和/或文本）特征注入到矩阵分解风格的排序目标中。例如，VBPR通过将视觉特征融入个性化排序扩展了BPR，而DeepStyle则探索利用深度视觉表示进行风格感知推荐。这些方法在概念上简单有效，但在很大程度上将多模态内容建模为静态辅助信息，且利用高阶用户-物品连通性的能力有限。

% 近期的工作已转向基于图的多模态建模，这主要受到图协同过滤（如LightGCN）中消息传递机制成功的启发。一个代表性的方向是构建多模态图神经网络推荐器，在交互图上传播信号，同时融合多视图（ID或内容）物品表示，例如MMGCN、DualGNN及多视图GCN的变体。除了直接在观察到的用户-物品图上应用消息传递外，一些研究还明确建模了源自多模态相似性的物品-物品关系。LATTICE引入了隐结构学习，用以构建和利用由多模态特征诱导的物品图，从而在共同交互之外丰富物品间的亲和力。然而，学习和更新此类结构可能会带来巨大的计算开销。FREEDOM通过将相对稳定的物品关系与噪声交互解耦，重新审视了这一设计：它冻结了静态物品图，专注于优化用户-物品交互图，提高了在大规模训练中的实用性。

% 鲁棒性也已成为多模态推荐的一个明确目标。MGCN利用行为信号过滤掉与偏好无关的分量，从而净化多模态特征，减少噪声或无信息内容的影响。BM3研究了多模态表示的自监督引导以增强泛化能力，而SMORE探索了用于模态融合和噪声抑制的频域建模，PGL则强调捕捉图学习中的高频局部结构特征。尽管取得了这些进展，但大多数现有的多模态推荐系统都隐含地假设多模态内容是可信的，且噪声主要源于特征或交互中的随机扰动。相比之下，现实世界的平台往往包含看似合理但具有误导性的多模态内容（如图像与文本不匹配、标题党描述）以及不可信的隐式交互，这些因素会产生系统性的偏差，其影响超出了随机扰动的范畴。这一差距促使我们从内容和交互两个维度研究可信多模态推荐：纠正多模态特征中的语义污染，并仔细考察在受损情况下，基于图的增强策略何时会放大不可信的协同信号。特别是，我们调查了由协同先验引导的图增强方法，并证明在协同信号受损时，这类方法反而可能产生不利影响。

% 噪声对应与多模态失配
% 除了特征扰动外，越来越多的多模态学习研究开始关注噪声对应关系，即跨模态的成对样本存在部分失配（伪阳性或伪阴性），这会严重损害跨模态对齐的效果。代表性方法通过估计软对应目标或实施一致性正则化来降低失配对的权重，例如BiCro和ESC，而鲁棒对比学习则在自监督目标中进一步缓解了噪声视图或伪阳性的影响。虽然这些技术主要针对对比对齐和检索目标而开发，但它们启发了一个与多模态推荐相关的关键原则：物品级的模态间对应关系可能是不可靠的，应当进行软纠偏，而不是将其视为完全准确的配对。遵循这一原则，我们设计了一种轻量级的基于对应关系的纠偏组件，它与成对排序目标兼容，并且可以以极小的开销插入到各种推荐系统中。

\section{Related Work}

\paragraph{Multimodal Recommendation}
Multimodal recommendation augments collaborative signals with item-side content such as images and texts to mitigate interaction sparsity and improve preference estimation\cite{vbpr,dvbpr,mrs_sv1,mrs_sv2}. Early multimodal recommenders typically injected visual (and/or textual) features extracted by pretrained encoders into matrix-factorization-style ranking objectives. For instance, VBPR~\cite{vbpr} extends BPR~\cite{bpr} by incorporating visual features into personalized ranking, while DeepStyle~\cite{deepstyle} explores leveraging deep visual representations for style-aware recommendation. These methods are conceptually simple and effective, yet they largely model multimodal content as static side information and have limited ability to exploit high-order user--item connectivity.

More recent work has shifted toward graph-based multimodal modeling, motivated by the success of message passing in graph collaborative filtering \cite{lightgcn,ngcf,pinsage}. A representative line builds multimodal graph neural recommenders that propagate signals on the interaction graph while fusing multi-view (ID/content) item representations, such as MMGCN~\cite{mmgcn}, DualGNN~\cite{dualgnn}, and Multi-View GCN variants \cite{mgcn,ego,dragon}. Beyond directly applying message passing on the observed user--item graph, several studies explicitly model item--item relations derived from multimodal similarity. LATTICE~\cite{lattice} introduces latent structure learning to construct and leverage item graphs induced by multimodal features, thereby enriching item affinity beyond co-interactions. However, learning and updating such structures can be computationally costly. FREEDOM~\cite{freedom} revisits this design by decoupling relatively stable item relations from noisy interactions: it freezes the (static) item graph and focuses on refining the user--item interaction graph, improving practicality for large-scale training\cite{damrs}.

Robustness has also become an explicit objective in multimodal recommendation. MGCN~\cite{mgcn} leverages behavioral signals to purify multimodal features by filtering preference-irrelevant components. BM3~\cite{bm3} studies self-supervised bootstrapping for better generalization, while SMORE~\cite{smore}, PGL~\cite{pgl}, and related methods focus on modality fusion and noise suppression from frequency-structure-aware perspectives. Despite these advances, most multimodal recommenders still implicitly assume that multimodal content is trustworthy and that noise mainly stems from stochastic perturbations. In practice, however, platforms often contain misleading-yet-plausible content (e.g., image--text mismatch, click-bait descriptions) and unreliable implicit interactions, which can systematically bias representation learning beyond random noise~\cite{graphda,grcn,denoise_imp}. This motivates studying trustworthiness from both content and interaction perspectives: rectifying semantic pollution in multimodal features, and examining when graph-based enhancement may amplify untrustworthy collaborative signals. In particular, we analyze collaborative-prior-guided graph enhancement (e.g., GraphDA~\cite{graphda}-like strategies) and show its effects can be non-monotonic under corrupted collaborative signals—helpful when priors align with true patterns but harmful when misaligned due to error amplification in message passing.

\paragraph{Noisy Correspondence and Multimodal Mismatch}
Beyond feature perturbations, a growing line of multimodal learning research studies noisy correspondence, where paired samples across modalities are partially mismatched (false positives/negatives), which can severely harm cross-modal alignment\cite{bicro, ncr,mvcln}.
Representative methods estimate soft correspondence targets or enforce consistency regularization to down-weight mismatched pairs, e.g., BiCro~\cite{bicro} and ESC~\cite{esc}, and robust contrastive learning further mitigates noisy views/false positives in self-supervised objectives~\cite{rcl_noisy_views}.
While these techniques are primarily developed for contrastive alignment and retrieval objectives, they motivate a key principle relevant to multimodal recommendation: correspondence between item-level modalities can be unreliable and should be softly rectified rather than treated as clean pairs.
Following this principle, we design a lightweight correspondence-based rectification that is compatible with pairwise ranking objectives and can be plugged into diverse recommenders with negligible overhead.

\section{Preliminaries}
\label{sec:prelim}

\paragraph{Setting and notation.}
We study implicit-feedback multimodal recommendation.
Let $\mathcal{U}=\{u_1,\dots,u_M\}$ and $\mathcal{I}=\{i_1,\dots,i_N\}$ denote the user and item sets, respectively.
Observed interactions form a binary matrix $\mathbf{Y}\in\{0,1\}^{M\times N}$, where $y_{ui}=1$ indicates an interaction (e.g., click/purchase).
Equivalently, we denote the observed positive edge set as $\mathcal{E}^{+}=\{(u,i)\mid y_{ui}=1\}$ and the user--item bipartite graph as $\mathcal{G}=(\mathcal{U}\cup\mathcal{I},\mathcal{E}^{+})$.
Each item $i$ is associated with pre-extracted multimodal features $\tilde{\mathbf{e}}_{i}^{m}\in\mathbb{R}^{d_m}$ for modality $m\in\mathcal{M}$.
In this work, $\mathcal{M}=\{v,t\}$ (image/text) with $d_v=4096$ and $d_t=384$.

\paragraph{Recommendation objective.}
Given $\mathcal{G}$ and $\{\tilde{\mathbf{e}}_{i}^{m}\}$, a multimodal recommender learns a scoring function $\hat{y}_{ui}$ for top-$K$ ranking.
All backbones considered in our experiments are optimized with the standard BPR objective with uniform negative sampling:
\begin{equation}
\mathcal{L}_{\text{BPR}}=\sum_{(u,i,j)} -\log \sigma\!\big(\hat{y}_{ui}-\hat{y}_{uj}\big) \;+\; \Omega_{\text{backbone}},
\end{equation}
where $(u,i)\in\mathcal{E}^{+}$ and $j$ is uniformly sampled from items not interacted by $u$, and $\Omega_{\text{backbone}}$ denotes backbone-specific regularizers/auxiliary losses.
In all experiments, we keep each backbone's original objective and tuned hyperparameters unchanged.

\paragraph{Trustworthiness-aware corruptions.}
We consider two practical sources of untrustworthiness and simulate them in controlled settings.

\textbf{(1) Modality misalignment (feature-level corruption).}
A fraction of item-side modality features can be semantically mismatched (e.g., wrong image/title).
To simulate this, for each modality $m\in\mathcal{M}$ and a corruption ratio $\eta_m\in[0,0.5]$, we uniformly sample a subset $\mathcal{S}_m\subseteq\mathcal{I}$ with $|\mathcal{S}_m|=\lfloor \eta_m |\mathcal{I}|\rfloor$ and randomly permute modality features within $\mathcal{S}_m$:
\begin{equation}
\{\tilde{\mathbf{e}}_{i}^{m}\}_{i\in\mathcal{S}_m}
\leftarrow
\operatorname{Permute}\!\left(\{\tilde{\mathbf{e}}_{i}^{m}\}_{i\in\mathcal{S}_m}\right).
\end{equation}
Image and text corruptions are applied independently, and all random seeds are fixed for reproducibility.
Unless stated otherwise, in this setting we treat the observed interaction graph $\mathcal{G}$ as reliable and focus on rectifying feature-level untrustworthiness.

\textbf{(2) Interaction noise (graph-level corruption).}
Observed interactions may contain spurious edges or miss meaningful ones.
We simulate graph-level corruption by randomly editing the training edges with a noise ratio $\eta_e$.
Specifically, edge deletion removes $\lfloor \eta_e|\mathcal{E}^{+}|\rfloor$ edges uniformly at random from $\mathcal{E}^{+}$, while edge addition uniformly samples the same number of non-existing user--item pairs $(u,i)\notin\mathcal{E}^{+}$ and adds them as positive training edges.
All random seeds are fixed.
Note that these add/delete operations are solely used to simulate interaction noise; our interaction-level analyses in Section~\ref{sec:insight} further study how injecting edges into the training set versus the propagation graph affects robustness under such noise.

\section{Methodology}
\label{sec:method}

\subsection{Backbone Multimodal Recommender}
\label{sec:backbone}

We consider a general multimodal recommendation setting with an observed user--item interaction graph $\mathcal{G}$ and item-side multimodal features $\{\tilde{\mathbf{e}}_i^{m}\}_{m\in\mathcal{M}}$ (e.g., image/text), where $\tilde{\mathbf{e}}_i^{m}\in\mathbb{R}^{d_m}$ denotes the observed feature of modality $m$ for item $i$.
A multimodal backbone predicts a preference score $\hat{y}_{ui}$ for each user--item pair, which we abstract as a scoring function
\begin{equation}
\hat{y}_{ui} = f_{\theta}\!\left(u, i;\, \mathcal{G},\, \{\tilde{\mathbf{e}}_i^{m}\}_{m\in\mathcal{M}}\right),
\end{equation}
where $\theta$ denotes backbone parameters.

\paragraph{Offline plug-and-play integration.}
Our Modality-level Rectification Module is applied offline as a preprocessing step: it takes the observed (potentially misaligned) modality features $\{\tilde{\mathbf{e}}_i^{m}\}$ and outputs rectified features $\{\mathbf{e}_{i}^{m,\text{rect}}\}$ with the same dimensionality as the original inputs.
We then fix $\{\mathbf{e}_{i}^{m,\text{rect}}\}$ and train the backbone with its original objective without architectural modification, by directly replacing the modality inputs:
\begin{equation}
f_{\theta}\!\left(u, i;\, \mathcal{G},\, \{\tilde{\mathbf{e}}_i^{m}\}\right)
\;\Rightarrow\;
f_{\theta}\!\left(u, i;\, \mathcal{G},\, \{\mathbf{e}_{i}^{m,\text{rect}}\}\right).
\end{equation}
In this work, we evaluate this integration on three representative multimodal recommenders, including LATTICE, FREEDOM, MGCN, SMORE and VBPR by only replacing their modality inputs with fixed ${\mathbf{e}_{i}^{m,\text{rect}}}$ while keeping architectures and objectives unchanged.

\subsection{Modality-level Rectification Module}
\label{sec:modal_rect}

In real-world e-commerce, the modality signal attached to an item may be misaligned with the item semantics (e.g., wrong images or misleading titles), which violates the one-to-one correspondence assumption between items and their multimodal content.
Following our trustworthiness-aware formulation, we model such feature-level untrustworthiness as an unreliable item--modality association (i.e., the observed modality feature of item $i$ may originate from another item).
To rectify misaligned modality features in a plug-and-play manner, we propose a \textbf{Modality-level Rectification Module} that learns soft correspondences between items and modality signals and then uses the correspondences to re-aggregate modality features.

\paragraph{Anchor embeddings.}
The key idea is to introduce a comparatively reliable anchor representation for each item and align multimodal signals to this anchor space.
In the modality-untrustworthy setting studied in this section, we assume the observed user--item interactions are reliable, while the multimodal features may be misaligned.
Accordingly, we obtain the anchor by pre-training a LightGCN encoder on the observed interaction graph and use its collaborative item embedding as the anchor:
\begin{equation}
\mathbf{e}_{i}=\frac{1}{L+1}\sum_{\ell=0}^{L}\mathbf{e}_{i}^{(\ell)}.
\end{equation}
We then $\ell_2$-normalize the anchor as $\bar{\mathbf{e}}_i=\mathbf{e}_i/\|\mathbf{e}_i\|$.

\paragraph{Step 1: Robust modality-to-anchor projection.}
Given the observed modality feature $\tilde{\mathbf{e}}_i^{m}\in\mathbb{R}^{d_m}$ for modality $m\in\mathcal{M}$, we learn a lightweight projection
\begin{equation}
\mathbf{z}_i^{m} = g_m(\tilde{\mathbf{e}}_i^{m}) \in \mathbb{R}^{d},
\end{equation}
where $g_m(\cdot)$ is implemented as a linear layer.
We normalize projected vectors as $\bar{\mathbf{z}}_i^{m}=\mathbf{z}_i^{m}/\|\mathbf{z}_i^{m}\|$ and adopt a cosine regression objective:
\begin{equation}
\ell_i^{m} = 1 - \langle \bar{\mathbf{e}}_i, \bar{\mathbf{z}}_i^{m}\rangle.
\end{equation}
Since a fraction of observed pairs $(i,\tilde{\mathbf{e}}_i^m)$ can be mismatched, directly minimizing the average loss is sensitive.
We thus employ a small-loss selection strategy: within each mini-batch, we keep only the lowest-loss (highest-similarity) subset and optimize their average:
\begin{equation}
\mathcal{L}_{\text{proj}}^{m} = \frac{1}{|\mathcal{B}_{\text{keep}}|}
\sum_{i\in\mathcal{B}_{\text{keep}}} \ell_i^{m},
\end{equation}
where $\mathcal{B}_{\text{keep}}$ keeps a fixed ratio $\rho$ of the smallest-loss instances in the batch.
This yields a robust projection that is less affected by mismatched item--modality pairs.

\paragraph{Step 2: Sparse affinity construction.}
After obtaining projected modality representations, we estimate soft correspondences between anchors and modality signals across all items.
We first compute the cosine similarity matrix
\begin{equation}
s_{ij}^{m}=\langle \bar{\mathbf{e}}_i, \bar{\mathbf{z}}_j^{m}\rangle,
\end{equation}
and then build a sparse affinity matrix $\mathbf{A}^{m}\in\mathbb{R}_{+}^{N\times N}$ by keeping only the top-$K$ neighbors for each anchor:
\begin{equation}
\mathbf{A}^{m}_{ij}=
\begin{cases}
\exp\!\big(s_{ij}^{m}/\tau\big), & j\in \operatorname{TopK}_{K}(\{s_{ij}^{m}\}_{j=1}^{N}),\\
0, & \text{otherwise},
\end{cases}
\end{equation}
where $\tau$ is a temperature hyper-parameter.
To avoid over-correction, we enforce that the diagonal entry $\mathbf{A}^{m}_{ii}$ is included in the sparse candidates.

\paragraph{Step 3: Sinkhorn-based soft matching.}
A row-normalized affinity may suffer from ``hubness'' (a few modality signals attract too many anchors).
To obtain a more balanced correspondence that approximates a soft one-to-one assignment, we apply a Sinkhorn--Knopp scaling procedure on $\mathbf{A}^{m}$ and obtain a soft matching matrix $\mathbf{P}^{m}$:
\begin{equation}
\mathbf{P}^{m}=\operatorname{diag}(\mathbf{u})\,\mathbf{A}^{m}\,\operatorname{diag}(\mathbf{v}),
\end{equation}
where $\mathbf{u},\mathbf{v}\in\mathbb{R}_{+}^{N}$ are iteratively updated to make $\mathbf{P}^{m}$ approximately doubly-stochastic (i.e., row/column sums close to constants):
\begin{equation}
\mathbf{u}\leftarrow \mathbf{1}\oslash(\mathbf{A}^{m}\mathbf{v}+\epsilon),\qquad
\mathbf{v}\leftarrow \mathbf{1}\oslash((\mathbf{A}^{m})^{\top}\mathbf{u}+\epsilon),
\end{equation}
with $\oslash$ denoting element-wise division.
We implement the above procedure on sparse matrices to keep the complexity linear in the number of non-zeros ($\mathcal{O}(NK)$) and we run Sinkhorn iterations with a small $\epsilon$ for numerical stability.

\paragraph{Step 4: Feature rectification via correspondence aggregation.}
Given the soft matching matrix $\mathbf{P}^{m}$, we rectify the raw modality features by aggregating signals from matched items:
\begin{equation}
\hat{\mathbf{e}}_{i}^{m}=\sum_{j=1}^{N}\mathbf{P}^{m}_{ij}\tilde{\mathbf{e}}_{j}^{m}.
\end{equation}
Finally, we adopt a diagonal prior mix to preserve part of the original observation and avoid aggressive replacement:
\begin{equation}
\mathbf{e}_{i}^{m,\text{rect}}=\lambda\,\tilde{\mathbf{e}}_{i}^{m}+(1-\lambda)\,\hat{\mathbf{e}}_{i}^{m},
\end{equation}
where $\lambda\in[0,1]$ controls the strength of rectification.

\section{Interaction-level Trustworthiness Insights}
\label{sec:insight}

Real-world implicit-feedback graphs are often untrustworthy: observed interactions may contain spurious edges (e.g., misclicks or exposure-driven behaviors) and miss meaningful ones.
A common practice to mitigate such noise is to edit the interaction graph using a collaborative prior (e.g., add ``high-confidence'' edges or prune ``low-confidence'' ones).
In this section, we examine this practice under controlled stress tests and show that similarity-based edge editing is not universally beneficial under noisy interactions: it can help or hurt depending on the noise type and the reliability (alignment) of the collaborative prior.

\subsection{Stress-test Setting}
We simulate untrustworthy interactions by randomly corrupting the training edge set $\mathcal{E}^{+}_{\text{tr}}$ with a noise ratio $\eta_e$ (Sec.~\ref{sec:prelim}): we either (i) delete a fraction of training edges, or (ii) add the same number of random non-existing user--item pairs as positive training edges.
This random add/delete is only used to construct noisy training data for stress testing; the edge editing strategy studied below is fully similarity-based and contains no random edge generation.

\subsection{Similarity-based Edge Editing with a Collaborative Prior}
\subsubsection{Collaborative prior and similarity}
Given a (possibly noisy) training graph, we pre-train a LightGCN encoder on the corrupted $\mathcal{E}^{+}_{\text{tr}}$ and obtain user/item embeddings $\mathbf{e}_u,\mathbf{e}_i$.
We define the collaborative similarity as
\begin{equation}
s_{ui}=\mathbf{e}_u^{\top}\mathbf{e}_i.
\end{equation}
We then study two widely-used graph editing operations separately: \textbf{edge pruning} and \textbf{relation completion}.
Importantly, both operations are rank-based and control the final number of edited edges with a fixed ratio $r$.
Specifically, we set $r=5\%$ to control the editing magnitude.
Pruning always removes exactly $r|\mathcal{E}^{+}_{\text{tr}}|$ edges, while completion attempts to add $r|\mathcal{E}^{+}_{\text{tr}}|$ candidate edges and may end up with fewer after holdout filtering.

\subsubsection{Editing operations}
\paragraph{Edge pruning (similarity-based deletion).}
To remove potentially spurious interactions, we delete the bottom-$r\%$ observed training edges according to $s_{ui}$:
\begin{equation}
\mathcal{E}^{\text{prune}}_{\text{tr}}
=\mathcal{E}^{+}_{\text{tr}}
\setminus
\operatorname{Bottom}_{r}\!\big(\{s_{ui}\mid (u,i)\in\mathcal{E}^{+}_{\text{tr}}\}\big),
\end{equation}
where $\operatorname{Bottom}_{r}(\cdot)$ selects the subset of observed edges whose similarities fall into the lowest $r\%$ among all training positives.

\paragraph{Relation completion (similarity-based addition).}
To recover potentially missing relations, we first generate high-confidence candidates from both user and item sides using top-$K$ selection:
\begin{align}
\mathcal{N}_{U}(u) &= \operatorname{TopK}_{K_{UI}^{(u)}}\big(\{i\in\mathcal{I}\mid (u,i)\notin\mathcal{E}^{+}_{\text{tr}}\},\, s_{ui}\big),\\
\mathcal{N}_{I}(i) &= \operatorname{TopK}_{K_{UI}^{(i)}}\big(\{u\in\mathcal{U}\mid (u,i)\notin\mathcal{E}^{+}_{\text{tr}}\},\, s_{ui}\big),
\end{align}
and take the union as the candidate set
\begin{equation}
\mathcal{C}_{\text{tr}}
=\big\{(u,i)\mid i\in \mathcal{N}_{U}(u)\big\}
\;\cup\;
\big\{(u,i)\mid u\in \mathcal{N}_{I}(i)\big\}.
\end{equation}
We then add the top-$r\%$ edges in $\mathcal{C}_{\text{tr}}$ by similarity, yielding
\begin{equation}
\mathcal{E}^{\text{comp}}_{\text{tr}}
=\mathcal{E}^{+}_{\text{tr}}
\;\cup\;
\operatorname{Top}_{r}\!\big(\{s_{ui}\mid (u,i)\in\mathcal{C}_{\text{tr}}\}\big),
\end{equation}
where $\operatorname{Top}_{r}(\cdot)$ selects the subset of candidate edges whose similarities fall into the highest $r\%$ among all candidates.
To avoid any information leakage, we further remove from the added edges those appearing in validation/test splits; therefore the final number of added edges is at most $r|\mathcal{E}^{+}_{\text{tr}}|$ and can be smaller, i.e.,
\begin{equation}
|\mathcal{E}^{\text{comp}}_{\text{tr}}|
\le (1+r)|\mathcal{E}^{+}_{\text{tr}}|.
\end{equation}

\subsubsection{Where to apply editing: supervision vs.\ propagation}
We further distinguish where to apply edge editing in graph-based recommenders.
Let $\mathcal{E}^{+}_{\text{tr}}$ denote the edge set used to form BPR training triplets (i.e., the supervision signal), and let $\mathcal{E}_{\text{prop}}$ denote the edge set used for message passing.
We consider three variants:
(\textsc{Train-only}) edit $\mathcal{E}^{+}_{\text{tr}}$ but keep $\mathcal{E}_{\text{prop}}$ unchanged;
(\textsc{Graph-only}) edit $\mathcal{E}_{\text{prop}}$ but keep $\mathcal{E}^{+}_{\text{tr}}$ unchanged;
(\textsc{Both}) edit both.
These variants share the same backbone and objective; only the edge sets used for supervision and/or propagation differ.

\subsection{Key Findings and Practical Takeaways}
\paragraph{Key findings and explanations.}
Across varying interaction-noise ratios, we observe non-monotonic effects of similarity-based completion and pruning (Sec.~\ref{sec:insight_exp}).
\textbf{When there is no interaction noise}, the LightGCN prior is generally reliable, and both completion and pruning are typically beneficial.
\textbf{When noise is injected}, the effectiveness of edge editing becomes unstable and can help or hurt depending on the noise type and the resulting prior--signal alignment.
In particular, under random edge additions, the learned similarity $s_{ui}$ is often contaminated, so the gains from completion/pruning often diminish and the base model can become competitive or even preferable.
Under random edge deletions, the collaborative prior can be sometimes useful and sometimes misleading: LightGCN may recover meaningful structure and benefit from completion/pruning in some cases, but may also mis-rank edges and degrade performance in others.
\textbf{Completion can hurt} because $s_{ui}$ learned from noisy interactions may be biased; adding ``high-similarity'' edges may introduce additional false positives into supervision and/or propagation.
\textbf{Pruning can hurt} because a noisy prior may mis-rank true preference edges, and deleting low-similarity edges may remove rare but informative signals in sparse graphs.
Moreover, \textsc{Graph-only} editing may amplify misaligned neighbors through multi-hop message passing, while \textsc{Train-only} editing directly perturbs supervision via label editing; applying edits to \textsc{Both} can compound these effects.
Overall, these results caution against treating collaborative-prior-based edge editing as a universal remedy for noisy implicit feedback, and motivate validating such edits under stress tests.

\paragraph{Practical takeaways.}
\textbf{(Evaluation caveat under filtering).}
In implicit-feedback evaluation, it is common to filter training positives from the candidate set when computing Recall/NDCG.
Therefore, \textsc{Train-only} and \textsc{Both} edge editing (label editing) changes the set of filtered items: deleting training positives reduces the filtered set and can make evaluation scores appear artificially higher, even when true preference learning does not improve.
To ensure fair comparison, filtering should be performed w.r.t.\ the original training positives before editing, or results should be reported under consistent candidate sets.

\textbf{(When denoising can backfire).}
Our results suggest that collaborative-prior-based edge editing is not a universal remedy under noisy interactions: editing supervision perturbs the learning signal, while editing the propagation graph can amplify erroneous neighbors through message passing.
In practice, edge editing should be validated under stress tests; when interaction trustworthiness is low or the prior is unstable, it can be safer to decouple editing from propagation or apply conservative edits.

\section{Experiments}

\subsection{Experimental Setup}

\subsubsection{Datasets}
\label{sec:datasets}
Following prior multimodal recommendation works, we conduct experiments on three Amazon Review subsets: \textit{Baby}, \textit{Sports and Outdoors}, and \textit{Electronics} (denoted as \textit{Baby}, \textit{Sports}, and \textit{Electronics}). We apply the standard 5-core preprocessing on both users and items, and report statistics in Table~\ref{tab:dataset_stats}. For the visual modality, we use 4,096-dimensional VGG16 features~\cite{VGG}. For the textual modality, we use sentence-transformers~\cite{sentence_trans} to encode the concatenation of brand, title, description, and category into a 384-dimensional embedding.

\begin{table}[t]
    \centering
    \caption{Dataset Statistics}
    \label{tab:dataset_stats}
    \begin{tabular}{lcccc}
        \toprule
        Dataset & \#User & \#Item & \#Interaction & Sparsity \\
        \midrule
        Baby     & 19,445 & 7,050  & 160,792 & 99.883\% \\
        Sports   & 35,598 & 18,357 & 296,337 & 99.955\% \\
        Electronics & 192,403 & 63,001 & 1,689,188 & 99.986\% \\
        \bottomrule
    \end{tabular}
\end{table}

\subsubsection{Backbones and Integration}
\label{sec:baselines}
We evaluate representative multimodal recommenders as backbones, covering both MF-style and graph-based paradigms: \textbf{VBPR}, \textbf{LATTICE}, \textbf{FREEDOM}, \textbf{SMORE}, and \textbf{MGCN}. Our modality-level rectification is an offline preprocessing step that replaces the original modality features $\{\tilde{\mathbf{e}}_i^{m}\}$ with rectified features $\{\mathbf{e}_{i}^{m,\text{rect}}\}$, while keeping each backbone's original architecture, objective, and tuned hyperparameters unchanged. We additionally study interaction-level edge editing under noisy interactions, which applies only to message-passing graph-based backbones.

\subsubsection{Evaluation Protocol}
\label{sec:evaluation}
We follow standard protocols in multimodal recommendation.
User--item interactions are split into 80\%/10\%/10\% for training/validation/testing.
Hyperparameters are tuned on validation and all results are reported on the test set.
We adopt full-ranking evaluation, where each user ranks all candidate items, and report Recall@$K$ and NDCG@$K$ averaged over test users. For all edge-editing variants, we perform filtering w.r.t.\ the original training positives (before editing) to ensure a consistent candidate set.

\subsubsection{Implementation Details}
\label{sec:implementation}
All methods are implemented in the MMrec framework~\cite{mmrec} with a fixed random seed.
We use batch size 2048 for training and 4096 for evaluation.
Models are trained for up to 1000 epochs with early stopping (patience 30 on validation Recall@10).
Parameters are initialized with Xavier and optimized with Adam (default settings unless specified).

The projection step adopts small-loss selection with keep ratio $\rho$.
In controlled modality-misalignment stress tests, we set $\rho=\eta_m+0.05$, where $\eta_m$ is the injected misalignment ratio and $0.05$ approximates inherent misalignment in raw data.
In practice, $\rho$ can be chosen via a lightweight estimate of the misalignment rate (e.g., random auditing of image--text pairs or simple consistency checks).

\begin{table*}[t]
\centering
\small
\setlength{\tabcolsep}{6pt} % 列间距
\renewcommand{\arraystretch}{1.15} % 行高

\caption{Clean performance on Baby, Sports, and Electronics. MR denotes our Modality-level Rectification applied as offline preprocessing by replacing modality inputs. We report Recall@10/20 and NDCG@10/20; the better result between the backbone and \textbf{+MR} is bolded. OOM indicates CUDA out-of-memory under this fixed setting.}
\label{tab:overall}

\begin{tabular}{
  l l
  S[table-format=1.4] S[table-format=1.4] % VBPR +MR
  S[table-format=1.4] S[table-format=1.4] % LATTICE +MR
  S[table-format=1.4] S[table-format=1.4] % MGCN +MR
  S[table-format=1.4] S[table-format=1.4] % FREEDOM +MR
  S[table-format=1.4] S[table-format=1.4]                      % SMORE +MR
}
\toprule
\multirow{2}{*}{Dataset} & \multirow{2}{*}{Metric}
& \multicolumn{2}{c}{MF-based model}
& \multicolumn{8}{c}{GNN-based model}
 \\
\cmidrule(lr){3-4}\cmidrule(lr){5-12}
& 
& {VBPR} & {\textbf{+MR}}
& {LATTICE} & {\textbf{+MR}} & {MGCN} & {\textbf{+MR}} & {FREEDOM} & {\textbf{+MR}} & {SMORE} & {\textbf{+MR}}
\\
\midrule

% ---------------- Baby ----------------
\multirow{4}{*}{Baby}
& R@10 & 0.0428 & 0.0428 & 0.0537 & \textbf{0.0562} & 0.0613 & \textbf{0.0627} & 0.0624 & \textbf{0.0641} & 0.0647 & \textbf{0.0684} \\
& R@20 & \textbf{0.0665} & 0.0658 & 0.0837 & \textbf{0.0852} & 0.0939 & \textbf{0.0946} & 0.0977 & \textbf{0.1001} & 0.1007 & \textbf{0.1037} \\
& N@10 & 0.0226 & \textbf{0.0231} & 0.0289 & \textbf{0.0301} & 0.0328 & \textbf{0.0343} & 0.0326 & \textbf{0.0341} & 0.0355 & \textbf{0.0379} \\
& N@20 & 0.0287 & \textbf{0.0291} & 0.0366 & \textbf{0.0376} & 0.0412 & \textbf{0.0425} & 0.0417 & \textbf{0.0433} & 0.0448 & \textbf{0.0469} \\
\midrule

% ---------------- Sports ----------------
\multirow{4}{*}{Sports}
& R@10 & 0.0550 & \textbf{0.0571} & \textbf{0.0625} & 0.0617 & \textbf{0.0713} & 0.0712 & 0.0717 & \textbf{0.0730} & 0.0736 & \textbf{0.0738} \\
& R@20 & 0.0841 & \textbf{0.0879} & \textbf{0.0954} & 0.0912 & \textbf{0.1060} & 0.1056 & 0.1088 & \textbf{0.1114} & 0.1105 & \textbf{0.1110} \\
& N@10 & 0.0300 & \textbf{0.0308} & \textbf{0.0337} & 0.0334 & 0.0385 & 0.0385 & 0.0389 & \textbf{0.0397} & \textbf{0.0405} & 0.0402 \\
& N@20 & 0.0375 & \textbf{0.0387} & \textbf{0.0423} & 0.0410 & \textbf{0.0475} & 0.0474 & 0.0485 & \textbf{0.0496} & \textbf{0.0500} & 0.0498 \\
\midrule

% ---------------- Electronics ----------------
\multirow{4}{*}{Electronics}
& R@10 & 0.0281 & \textbf{0.0299} & OOM & OOM & 0.0418 & 0.0418 & 0.0377 & \textbf{0.0397} & 0.0400 & \textbf{0.0437} \\
& R@20 & 0.0434 & \textbf{0.0465} & OOM & OOM & 0.0619 & 0.0619 & 0.0577 & \textbf{0.0602} & 0.0648 & \textbf{0.0650} \\
& N@10 & 0.0152 & \textbf{0.0161} & OOM & OOM & 0.0234 & 0.0234 & 0.0207 & \textbf{0.0218} & \textbf{0.0247} & 0.0244 \\
& N@20 & 0.0192 & \textbf{0.0204} & OOM & OOM & 0.0287 & 0.0287 & 0.0259 & \textbf{0.0271} & \textbf{0.0301} & 0.0299 \\
\bottomrule
\end{tabular}
\end{table*}

\subsection{Modality-level Rectification: Clean Performance}
\label{sec:exp_clean}

Table~\ref{tab:overall} reports the clean-setting performance of five multimodal recommenders (VBPR, LATTICE, MGCN, FREEDOM, and SMORE) on three datasets (Baby, Sports, and Electronics).
We denote our Modality-level Rectification as \textbf{MR} and integrate it offline by directly replacing the original modality features with rectified ones while keeping each backbone's architecture, objective (including backbone-specific losses/regularizers), and tuned hyperparameters unchanged.
For each metric, we bold the better result between the backbone and its \textbf{+MR} variant.

\paragraph{MR is non-invasive on clean inputs and often yields consistent gains.}
Overall, adding \textbf{MR} does not degrade clean performance and frequently brings noticeable improvements across datasets and backbones.
On Baby, \textbf{+MR} improves all GNN-based backbones (LATTICE/MGCN/FREEDOM/SMORE) consistently on both Recall and NDCG metrics, while VBPR remains essentially unchanged on R@10 and shows a small gain on NDCG.
On Sports, \textbf{+MR} improves VBPR and the stronger GNN-based backbones (FREEDOM/SMORE) in most metrics, with only marginal differences for some model--metric pairs (e.g., LATTICE and MGCN).
On Electronics, \textbf{+MR} again yields consistent improvements for VBPR, FREEDOM, and SMORE across all four metrics, while MGCN is largely unchanged.

\paragraph{Discussion: why can MR help even without corruption?}
Although MR is designed for modality misalignment, the clean-setting gains suggest that real-world multimodal features may still contain mild noise or semantic inconsistency.
By learning soft item--modality correspondences and applying a conservative diagonal-prior mix, MR can suppress subtle mismatches while preserving the original signals, making it a safe preprocessing step for downstream recommenders.

\paragraph{Practical note on memory.}
For LATTICE on Electronics, we observe CUDA out-of-memory (OOM) under our fixed experimental setting when keeping the backbone and hyperparameters unchanged.
We therefore omit these results and focus on the remaining backbones, where MR can be applied efficiently and consistently.

\subsection{Modality-level Rectification: Robustness Under Modality Corruption}
\label{sec:exp_modal_robust}

\begin{figure}[t]
    \centering

    \includegraphics[width=\linewidth]{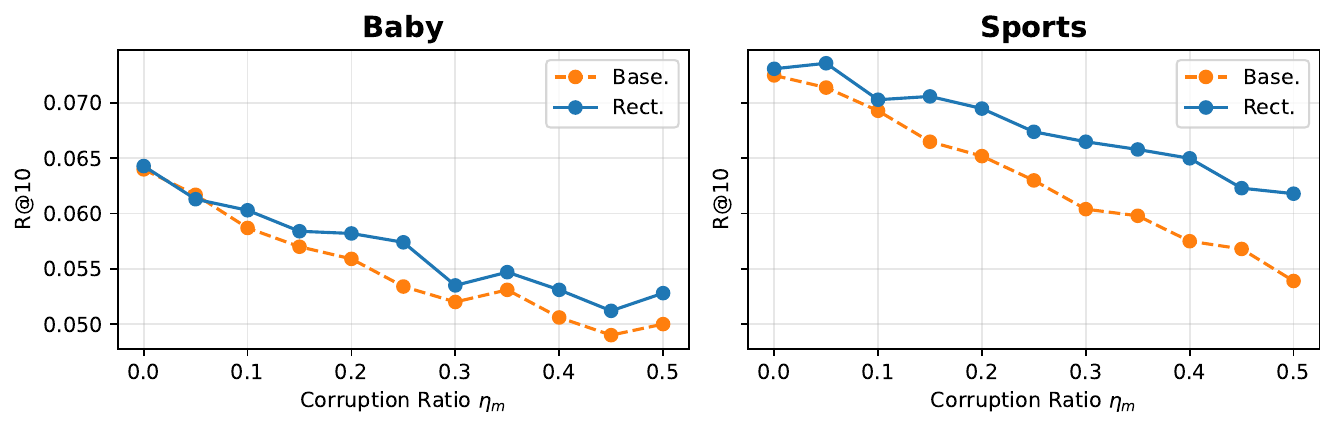}
     {\small (a) FREEDOM}

    \vspace{2mm}

    \includegraphics[width=\linewidth]{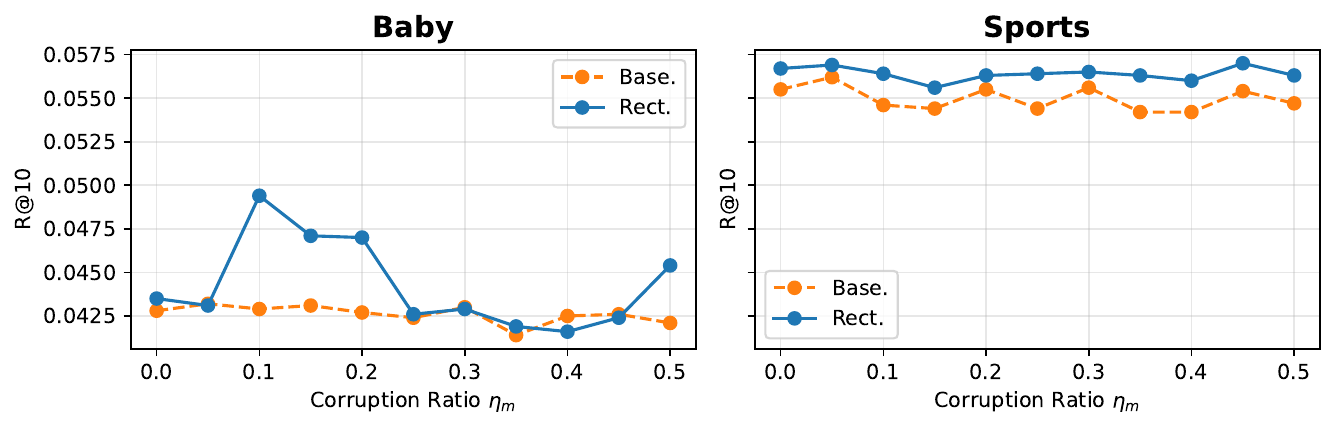}
     {\small (b) VBPR}

    \caption{Robustness under modality misalignment. Recall@10 of FREEDOM and VBPR on Baby and Sports as the modality corruption ratio $\eta_m$ increases. \textsc{Base.} denotes the original backbone using corrupted features, and \textsc{Rect.} applies our Modality-level Rectification as offline preprocessing.}

    \label{fig:stress_test}
\end{figure}

We evaluate the robustness of the proposed Modality-level Rectification under feature-level untrustworthiness by progressively increasing the modality-misalignment ratio $\eta_m$ from $0$ to $0.5$.
Following the corruption protocol in Sec.~\ref{sec:prelim}, we independently permute a fraction of image/text features and report Recall@10 on two datasets (Baby and Sports).
Figure~\ref{fig:stress_test} compares the vanilla backbones (\textsc{Base.}) and their rectified counterparts (\textsc{Rect.}) on a graph-based multimodal model (FREEDOM) and an MF-style model (VBPR).

\paragraph{Overall robustness gains across model families.}
As shown in Figure~\ref{fig:stress_test}, rectification consistently improves robustness on both the graph-based backbone (FREEDOM) and the MF-style backbone (VBPR) across datasets and noise levels.
The improvement becomes more evident as $\eta_m$ increases, indicating that learning soft item--modality correspondences effectively mitigates the impact of modality misalignment on downstream preference learning.

\paragraph{Why FREEDOM benefits more under modality mismatch.}
FREEDOM exhibits a clear performance degradation as $\eta_m$ grows, especially on Sports, whereas the rectified version maintains substantially higher Recall@10.
This is expected because FREEDOM leverages modality features to construct modality-induced item--item relations; once modality features are mismatched, the induced item--item graph can introduce erroneous neighbors and amplify unreliable signals through message passing, leading to sharper performance drops.
By rectifying item-side modality features before they are used to form such relations, our module reduces the chance of constructing misleading item--item connections, hence improving robustness.

\paragraph{VBPR is intrinsically more robust, yet still benefits from rectification.}
In contrast, VBPR is relatively insensitive to modality misalignment: its performance curves are much flatter across $\eta_m$.
A plausible reason is that VBPR incorporates multimodal features as an additive preference term in an MF-style scoring function; when modality information becomes unreliable, the model can rely more on the collaborative (ID) component, thus maintaining performance.
Nevertheless, rectification still brings consistent gains, and on Baby we observe a notable improvement around $\eta_m\!\approx\!0.1$--$0.2$.
We conjecture that the soft rectification can partially absorb corrupted modality signals, making the residual modality input less misleading and yielding an effect akin to ``noise injection + denoised reconstruction'', which helps VBPR extract more useful content information in this regime.

Overall, these results demonstrate that the proposed rectification serves as a robust, plug-and-play preprocessing module that improves multimodal recommendation under modality misalignment, benefiting both graph-based and MF-style backbones.

\begin{figure}[t]
    \centering

    \includegraphics[width=\linewidth]{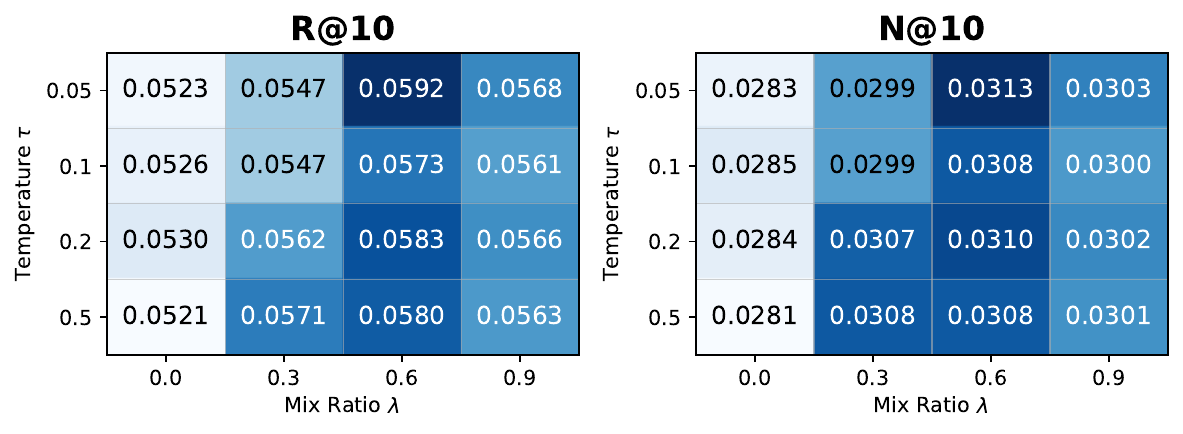}
     {\small (a) Baby}

    \vspace{2mm}

    \includegraphics[width=\linewidth]{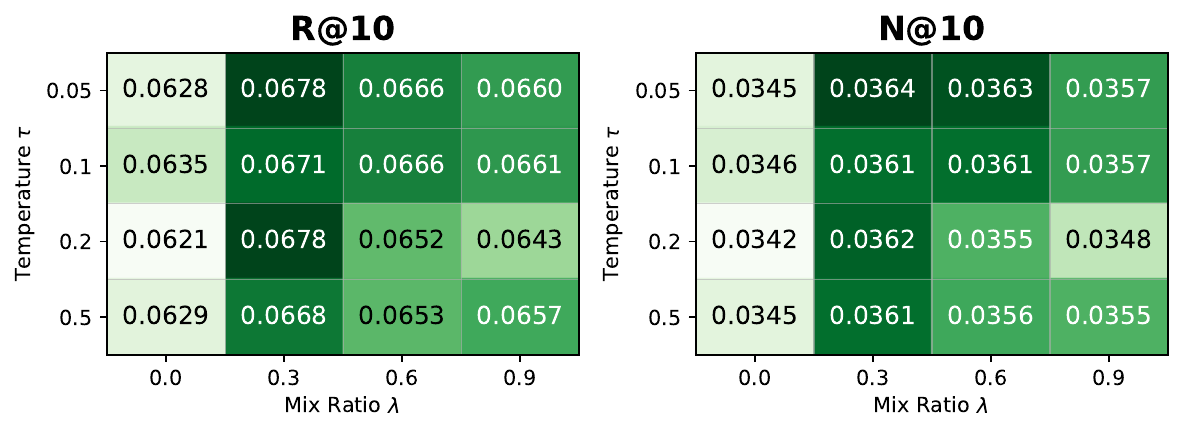}
     {\small (b) Sports}

    \caption{Hyperparameter sensitivity of MR under modality corruption ($\eta_m=0.2$) on FREEDOM. We vary the mix ratio $\lambda$ and temperature $\tau$, and report Recall@10 and NDCG@10 on Baby and Sports.}

    \label{fig:hyper_test}
\end{figure}
\subsection{Hyperparameter Sensitivity}
\label{sec:exp_hyper}

We study the sensitivity of Modality-level Rectification to two key hyperparameters: the diagonal prior mix ratio $\lambda$ in Eq.~(13) and the temperature $\tau$ in the affinity construction (Eq.~(10)).
To highlight the effect under feature-level untrustworthiness, we conduct experiments with a fixed modality corruption ratio $\eta_m=20\%$ and evaluate FREEDOM on Baby and Sports using Recall@10 and NDCG@10.

Figure~\ref{fig:hyper_test} shows that intermediate mixing generally yields the best performance.
Across both datasets and metrics, the sweet spot lies around $\lambda\approx0.6$, suggesting that a conservative rectification is beneficial: fully replacing the observed features ($\lambda=0$) can be too aggressive, while relying almost entirely on the corrupted observation ($\lambda\rightarrow 1$) weakens the correction effect.
Overall, MR is relatively robust to $\lambda$ within a broad mid-range (roughly $0.3$--$0.6$), while extreme values tend to be suboptimal.

The temperature $\tau$ controls the sharpness of the sparse affinity weights before Sinkhorn scaling.
We observe that MR is stable across a reasonable range of $\tau$, with $\tau$ in the lower-to-mid regime (e.g., $0.05$--$0.2$) typically performing well.
Very large temperatures (e.g., $\tau=0.5$) make affinities overly smooth, which may dilute the correspondence signal and slightly reduce performance.

In summary, MR is stable w.r.t.\ both hyperparameters under modality corruption; we use $\lambda=0.5$ as a stable default within the broad optimal region and set $\tau=0.1$ in all experiments. We further provide an ablation study in Appendix~\ref{app:ablation_mr}, verifying the contribution of Sinkhorn balancing and small-loss projection.

\begin{figure}[t]
    \centering
    \includegraphics[width=\linewidth]{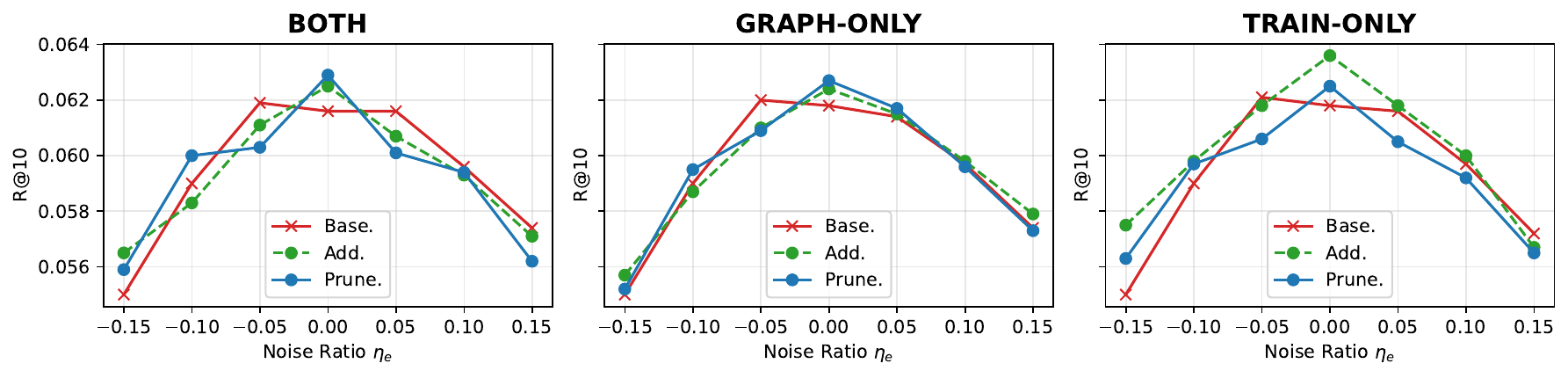}
    \caption{Interaction-level stress test on Baby with MGCN. We report Recall@10 as the interaction noise ratio $\eta_e$ varies. \textsc{Base.} is the original backbone; \textsc{Add.} and \textsc{Prune.} apply similarity-based relation completion or edge pruning using a LightGCN prior. \textsc{Train-only} edits BPR positives, \textsc{Graph-only} edits the propagation graph, and \textsc{Both} edits both supervision and propagation.}
    \label{fig:insight}
\end{figure}

\subsection{Interaction-level Insight Results}
\label{sec:insight_exp}

We empirically validate our interaction-level insights by stress-testing a collaborative-prior-based edge editing pipeline on a graph-based multimodal recommender.
Specifically, we report Recall@10 of \textsc{MGCN} on the Baby dataset under interaction noise ratios $\eta_e\in[-0.15,0.15]$ (Sec.~\ref{sec:prelim}), where negative values correspond to random edge deletions and positive values correspond to random edge additions in the training interactions.
We compare the vanilla backbone (\textsc{Base.}) with two similarity-based editing operations derived from a LightGCN prior: \textsc{Add.} (relation completion) and \textsc{Prune.} (edge pruning).
We further evaluate three ways to apply the edits: \textsc{Train-only} (edit BPR positives), \textsc{Graph-only} (edit the propagation graph), and \textsc{Both} (edit both supervision and propagation).

\paragraph{No-noise regime: prior-based completion/pruning is beneficial.}
As shown in Figure~\ref{fig:insight}, when there is no interaction noise ($\eta_e=0$), both \textsc{Add.} and \textsc{Prune.} consistently outperform the base model across all three application choices.
This indicates that, when the observed training interactions are reliable, the LightGCN prior provides a meaningful similarity signal $s_{ui}$ and similarity-based edge editing can refine neighborhoods and improve learning.

\paragraph{Under stress tests: the effect becomes unstable and can backfire.}
Once interaction noise is injected, the benefit of edge editing becomes non-monotonic and depends on the noise type and the alignment between the learned prior and the true preference signal.
Under random edge additions ($\eta_e>0$), the collaborative prior is more likely to be contaminated by spurious positives; as a result, the gains of \textsc{Add.}/\textsc{Prune.} diminish and can even disappear, making the base model competitive or preferable in some settings.
Under random edge deletions ($\eta_e<0$), the prior can sometimes help recover useful structure, but can also become misleading when missing edges distort neighborhood information, leading to mixed outcomes.

\paragraph{Train-only vs.\ graph-only vs.\ both.}
Across noise levels, \textsc{Train-only} editing tends to be more volatile since it directly performs label editing on BPR positives and thus perturbs the supervision signal.
\textsc{Graph-only} editing isolates the intervention to message passing and can be more stable, yet it may still amplify misaligned neighbors when the prior is noisy.
Applying edits to \textsc{Both} can compound these effects: it simultaneously changes the supervision signal and the propagation neighborhood, which can magnify the impact of prior contamination under interaction noise.

Overall, Figure~\ref{fig:insight} confirms that collaborative-prior-based edge editing is not a universal remedy for noisy implicit feedback: it is reliably beneficial in clean graphs but can become unstable and even harmful under stress tests, motivating careful validation and conservative usage in practice.

% 结论

% 在这项工作中，我们从内容和交互两个维度研究了多模态推荐系统（MRS）中的可信性挑战。我们提出了一种即插即用的模态级纠偏组件，通过轻量级投影和基于 Sinkhorn 的软匹配来减轻不可信模态特征（如视觉不一致的图像或误导性文本）的影响，在抑制失配模态信号的同时保留了语义一致的信息。

% 我们进一步在受控压力测试下研究了交互级的可信性，发现基于协同先验的图编辑（包括基于相似度的补全和剪枝）并非在所有情况下都有益。当交互数据干净时，此类编辑通常是有帮助的；然而，在交互受损的情况下，其效果变得非单调，即可能提升也可能损害性能，具体取决于噪声类型和协同先验的可靠性（对齐程度）。此外，对传播图进行编辑可能会通过消息传递放大失配的关系，而直接编辑监督信号则会干扰学习信号，这要求在部署前进行仔细的验证。

% 我们的研究目前局限于图像/文本模态和特定的受损模式，未来的工作可以扩展到更多模态，以及内容和交互中更真实的可信性失效场景。此外，在不可信交互下设计有原则的、感知对齐的图编辑与传播机制仍然是一个重要的研究方向。

% 在多个真实数据集和骨干模型上进行的大量实验证明，模态纠偏组件提高了系统的鲁棒性，并验证了上述交互级的观察结果。所有代码和数据均已公开，以供进一步的研究和实验。
\section{Conclusion}

This work studies trustworthiness in multimodal recommender systems (MRSs) from both content and interaction perspectives. We propose a plug-and-play modality-level rectification module that mitigates untrustworthy multimodal features (e.g., mismatched images or misleading text) via lightweight projection and Sinkhorn-based soft matching, suppressing mismatched signals while preserving consistent semantics.

We further conduct controlled stress tests on interaction-level trustworthiness and show that collaborative-prior-based graph editing (similarity-based completion and pruning) is not universally beneficial: it is typically helpful when interactions are clean, but becomes non-monotonic under interaction corruption, helping or hurting depending on noise type and the alignment (reliability) of the learned prior. Moreover, editing the propagation graph may amplify misaligned relations via message passing, while editing supervision perturbs the learning signal, highlighting the need for careful validation before deployment.

Our study focuses on image/text modalities and specific corruption patterns; future work may extend to additional modalities and more realistic trustworthiness failures. Designing principled, alignment-aware mechanisms for graph editing and propagation under untrustworthy interactions also remains an important direction. Extensive experiments across datasets and backbones demonstrate the robustness gains from rectification and support our interaction-level findings, and we will release code and data to facilitate further research.

% \begin{acks}
% % To Robert, for the bagels and explaining CMYK and color spaces.
% \end{acks}

\bibliographystyle{ACM-Reference-Format}
\bibliography{software}

% %%
% %% If your work has an appendix, this is the place to put it.
\appendix

\begin{figure}[t]
    \centering
    \includegraphics[width=\linewidth]{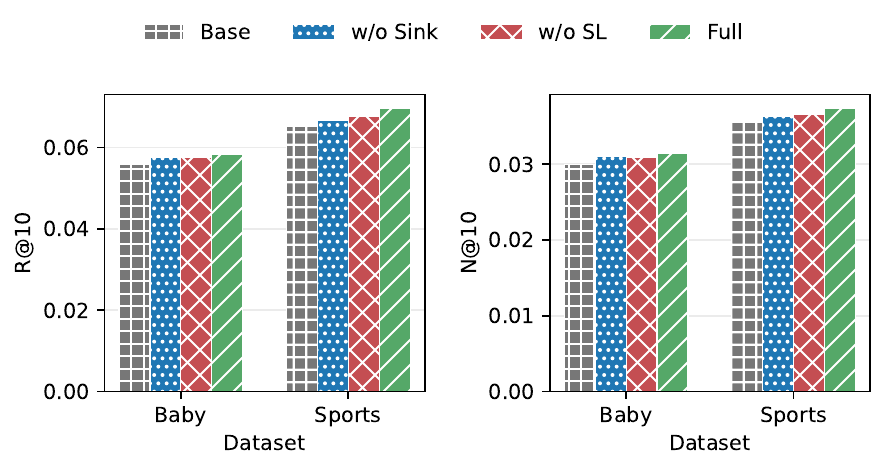}
    \caption{Ablation study of MR under modality misalignment ($\eta_m=20\%$) with FREEDOM on Baby and Sports. We report Recall@10 and NDCG@10. \textsc{Base}: w/o MR; \textsc{Full}: full MR; \textsc{w/o Sink}: replace Sinkhorn matching with row-normalized affinity; \textsc{w/o SL}: remove small-loss selection in projection training.}
    \label{fig:ablation}
\end{figure}

\section{Modality-level Rectification: Ablation Study}
\label{app:ablation_mr}

We provide a lightweight ablation study to verify the necessity of key components in Modality-level Rectification (MR).
Following the modality-corruption protocol in Sec.~\ref{sec:prelim}, we set the misalignment ratio to $\eta_m=20\%$ and evaluate \textsc{FREEDOM} on Baby and Sports using Recall@10 and NDCG@10.

\paragraph{Compared variants.}
We consider four variants:
(\textsc{Base}) w/o MR, directly using the corrupted modality features;
(\textsc{Full}) the complete MR module;
(\textsc{w/o Sink}) replacing Sinkhorn-based balancing with simple row normalization (row-norm) on the sparse affinity;
(\textsc{w/o SL}) removing the small-loss selection strategy and training the projection with the full-batch loss.

\paragraph{Results and analysis.}
As shown in Figure~\ref{fig:ablation}, \textsc{Full} achieves the best performance on both datasets and both metrics, confirming that MR is effective under modality misalignment.
Both \textsc{w/o Sink} and \textsc{w/o SL} lead to consistent drops compared with \textsc{Full}, indicating that (i) Sinkhorn balancing is helpful for mitigating hubness and producing more stable correspondences, and (ii) robust projection with small-loss selection improves alignment under mismatched item--modality pairs.
Finally, \textsc{Base} performs the worst, highlighting the severity of modality corruption and the necessity of rectification.
Overall, the ablation supports that the proposed design components are complementary and jointly contribute to robustness.


% % Lorem ipsum dolor sit amet, consectetur adipiscing elit. Morbi
% % malesuada, quam in pulvinar varius, metus nunc fermentum urna, id
% % sollicitudin purus odio sit amet enim. Aliquam ullamcorper eu ipsum
% % vel mollis. Curabitur quis dictum nisl. Phasellus vel semper risus, et
% % lacinia dolor. Integer ultricies commodo sem nec semper.


% % Etiam commodo feugiat nisl pulvinar pellentesque. Etiam auctor sodales
% % ligula, non varius nibh pulvinar semper. Suspendisse nec lectus non
% % ipsum convallis congue hendrerit vitae sapien. Donec at laoreet
% % eros. Vivamus non purus placerat, scelerisque diam eu, cursus
% % ante. Etiam aliquam tortor auctor efficitur mattis.


% % Nam id fermentum dui. Suspendisse sagittis tortor a nulla mollis, in
% % pulvinar ex pretium. Sed interdum orci quis metus euismod, et sagittis
% % enim maximus. Vestibulum gravida massa ut felis suscipit
% % congue. Quisque mattis elit a risus ultrices commodo venenatis eget
% % dui. Etiam sagittis eleifend elementum.

% % Nam interdum magna at lectus dignissim, ac dignissim lorem
% % rhoncus. Maecenas eu arcu ac neque placerat aliquam. Nunc pulvinar
% % massa et mattis lacinia.

\end{document}